\documentstyle[aasms4,amssym,psfig,epsfig,11pt]{article}

\def\d{{\rm\,d}} 
\def\K{{\rm\,K}}
\def\cm{{\rm\,cm}}

\lefthead{Liu, Fromerth, and Melia}
\righthead{Line Emission from M31*}

\begin{document}
\centerline{Submitted to the Astrophysical Journal}
\bigskip
\title{Line Emission from Cooling Accretion Flows in the Nucleus of M31}

\author{Siming Liu}
\affil{Department of Physics, The University of Arizona, Tucson, AZ 85721}

\author{Michael J. Fromerth\altaffilmark{1}}
\affil{Department of Physics, The University of Arizona, Tucson, AZ 85721}

\and

\author{Fulvio Melia\altaffilmark{2}}
\affil{Department of Physics and Steward Observatory, The University of Arizona, Tucson, 
AZ 85721}
\altaffiltext{1}{NSF Graduate Fellow.}
\altaffiltext{2}{Sir Thomas Lyle Fellow and Miegunyah Fellow.}

\begin{abstract}

The recent {\it Chandra} X-ray observations of the nucleus of M31, combined 
with earlier VLA radio and HST UV spectral measurements, provide the strictest 
constraints thus far available on the nature of accretion onto the supermassive 
black hole (called M31* hereafter) in this region. One of the two newly-detected
sources within roughly an arcsec of M31* may be its X-ray counterpart. If not,
the X-ray flux from the nucleus must be even lower than inferred previously.  
Some uncertainty remains regarding the origin of the UV excess from the compact 
component known as P2. In our earlier analysis, we developed a unified picture 
for the broadband spectrum of this source and concluded that M31* could be 
understood on the basis of the accretion model for Sgr A* at the Galactic 
center, though with several crucial differences.  Contrary to the `standard' 
picture in which the infalling plasma attains temperatures in excess of 
$10^{10}$ K near the event horizon, the best fit model for M31*, under the 
assumption that the UV radiation is in fact produced by this source, appears 
to correspond to a cool branch solution, arising from strong line cooling 
inside the capture radius. Starting its infall with a temperature of about 
$10^6$ K in the post-shock region, the plasma cools down efficiently to about 
$10^4$ K toward smaller radii. An important prediction of 
this model is the appearance of a prominent UV spike from hydrogen line emission,
which for simplicity was handled only crudely in the earlier work. It is our 
purpose here to model this line emission with significantly greater accuracy, 
using the algorithm CLOUDY, and to correctly take into account the attenuation 
along the line of sight.  We show that this level of sophistication reproduces 
the currently available multi-wavelength observations very well.  Very importantly, 
we predict a spectrum with several additional prominent emission lines that can be 
used to verify the model with future high-resolution observations.  A non-detection
of the predicted line emission from M31* would then tilt the favored
accretion picture in the direction of a hot Sgr A*-type of model, though with
only a single point remaining in the spectrum of M31*, additional observations
at other wavelengths would be required to seriously constrain this system.
\end{abstract}


\section{INTRODUCTION}

The nucleus of M31 contains two components, P1 \& P2, separated by about
$0\farcs5\approx 1.9$ pc (Lauer et al. 1993). Its optical image is
dominated by P1 while there is a UV upturn in P2 (King et al. 1995). The kinematics 
of stars near the nucleus (Kormendy \& Bender 1999) suggests that P1 \& P2 may be
identified as the turning points of an elliptical distribution of stars orbiting 
the central black hole of mass $3.0\times 10^7\;M_\odot$ in a slightly
eccentric orbit (Tremaine 1995). However, the relatively intense UV emission from 
P2 remains a puzzle.

Based on earlier radio observations of M31* (Crane, Dickel, \& Cowan 1992), Melia (1992)
suggested that its nature may be similar to that of the supermassive black hole, Sgr A*, at
the Galactic center. Identifying the X-ray object TF56 (Trinchieri \& Fabbiano 1991) with
M31*, King et al. (1995) proposed a compact nonthermal source at this location. However,
subsequent high resolution HST observations have partially resolved P2 (Lauer et al. 1998;
Brown et al. 1998), showing that it has a half-power radius of $0\farcs06$. Lauer et al.
(1998) argued that these new observations point to the UV upturn as being produced by a
tightly bound cluster of early type stars.  Even so, this picture does not appear to be
complete.

The apparent identification of an X-ray source with M31* (Garcia et al. 2000)
suggested that the accretion process in M31* and Sgr A* must be quite different
(Liu \& Melia 2001). If neither of the two sources detected within about
an arcsecond of M31* are its counterpart, this would place a strict upper
limit on its X-ray flux and hence the temperature of the emitting gas. 
In this picture, the UV radiation is an extension of M31*'s 
radio component, which, however, must turn over below $\sim 1$ keV. 

Instead of the model for Sgr A* in which the accreting plasma reaches temperatures 
of $10^{10}$ K or higher (e.g., Shapiro 1973a; Melia 1992, 1994), M31* appears to 
correspond well to a second branch of solutions in which cooling dominates heating 
and the gas settles down to a much lower temperature during its infall.  An optically 
thin plasma with a temperature of about $10^5-10^7$ K, a typical value at the capture 
radius $r_C\equiv 2GM/v_\infty^2$ of a supermassive black hole accreting from
stellar winds (Coker \& Melia 2000), lies on the unstable portion of the cooling curve
(Gehrels \& Williams 1993).  Here, $v_\infty$ is the velocity of the ambient gas 
flowing past the central black hole.  Liu \& Melia (2001) showed that in this case 
two branches of solutions exist, distinguished by the relative importance of cooling 
versus compressional heating at $r_C$. Depending on the initial temperature 
$T(r_C)$ and the mass accretion rate $\dot{M}$, the plasma either settles onto a hot 
branch or a cool branch, characterized by temperatures of $\sim 10^{10}$~K and 
$\sim 10^4$~K, respectively, toward smaller radii.

Comparing these two solutions to the observations, Liu \& Melia (2001) argued that 
M31* must lie on the cool branch, and that we should therefore see a spectral 
signature of its cooling flow, particularly a prominent UV spike due to hydrogen 
line emission and soft X-ray line recombination.  The purpose of this paper is to 
present a significantly more detailed calculation of the broadband emission spectrum, 
correcting also for the attenuation along the line of sight. In section 
\ref{accretion}, we discuss the standard Bondi-Hoyle model for supermassive black 
holes accreting from the ambient medium. In section \ref{lines} \& \ref{results}, 
we discuss the specific details of our line emission calculations and we identify 
several emission lines that may be used to verify our model.  Section \ref{discuss} 
presents the conclusions.

\section{LOW ANGULAR MOMENTUM ACCRETION FLOW}
\label{accretion}

In this section, we briefly summarize the Bondi-Hoyle accretion model for a low luminosity
accreting supermassive black hole embedded within a gaseous environment. The latter may be
produced by the winds of nearby stars.  Gas flowing past the black hole will be captured
at a rate determined by its velocity and specific angular momentum relative to the compact
object. Because winds far from the stars are always highly supersonic, many shocks can form
where the winds collide before the gas is captured within the central gravitational field.
These shocks dissipate most of the kinetic energy and heat the plasma to a temperature of 
$\sim (3\ v_\infty /4)^2\mu/3 R_g$, where $R_g$ is the gas constant and $\mu$ is the 
molecular weight per particle, which is about $1/2$ for a fully ionized plasma. 
During the infall, the plasma accelerates under the influence of gravity and again
goes transonic.  We base our analysis on the global structure inferred for these
flows from existing 3D hydrodynamical simulations of the infalling gas (e.g., Coker \& 
Melia 1997), and we take our starting point to be a radius where the plasma has already 
re-crossed into the supersonic region.  Our analysis will adopt a purely radial flow,
for which consistency requires a negligibly small specific angular momentum in the 
accreting gas. 

Due to their low luminosity, supermassive black holes accreting in this fashion can be
treated with significant simplification.  For example, the energy loss due to radiation
can be treated with just a cooling term in the energy equation and the related effect 
on the dynamics of the flow is negligible. In this sense, the structure of the accretion
flow and the radiation may be handled separately.  

\subsection{Dynamical Equations}
The stress-energy tensor for a radial accretion flow with a frozen-in magnetic field
is given by (Novikov \& Thorne, 1973):
\begin{eqnarray}
T^{\delta\beta}&=& \epsilon\ u^\delta u^\beta + p\ (g^{\delta\beta}+u^\delta u^\beta) +
{1\over 8\pi}(2 B^2 u^\delta u^\beta + B^2 g^{\delta\beta} - 2 B^\delta B^\beta) \ \ ,  
\end{eqnarray}
where $\epsilon = \alpha n k_b T/ 2 \mu + n m_p c^2$ is the energy density. The
general expression for $\alpha$ is given by Chandrasekhar (1939):
\begin{equation}
\alpha = x\left({3K_3(x)+K_1(x)\over 4K_2(x)}-1\right) + y\left({3K_3(y)+K_1(y)\over
4K_2(y)}-1\right) \ \  ,
\end{equation}
where $x=m_e c^2/k_b T$, $y=m_p c^2/k_b T$ and $K_i$ is the ith order modified Bessel
function. Also, $n$ is the baryon number density, and $p = n k_b T/\mu$ is the gas pressure. The
other symbols have their usual meaning. The shear and bulk viscosity has been neglected
in the equation because the flow is spherically symmetric. In the case that the
radiation is not very strong, the radiation can be treated as an external field,
which just contributes a cooling term to the accretion flow. So we don't include it in
the stress-energy tensor either.

In Schwarszchild coordinates, the metric is given as
\begin{equation}
\d s^2=-c^2\d\tau^2=-(1-{2GM\over r\ c^2})\ c^2\d t^2+{1\over 1-{2GM\over
r\ c^2}}\d r^2+r^2(\d\theta^2+\sin^2{\theta}\d\phi^2)\ \ ,
\end{equation}
where $M$ is the mass of the supermassive black hole. Then we have 
$u^\alpha=(u^t, u^r, 0, 0) = (\d t/\d\tau, \d r/c\d\tau, 0, 0)$, which is dimensionless in
our notation. Because the magnetic field in a spherical accretion flow is dominated by its 
radial component (Shapiro 1973b), one has $B^\alpha = (B^t, B^r, 0, 0)$. The electric
field is always zero in the comoving frame of a fully ionized plasma with frozen-in
magnetic field, so ${\bf B}$ is orthogonal to ${\bf u}$:
\begin{equation}
B^\alpha u_\alpha =0\ .
\end{equation}

The dynamical equations for the radial accretion flow are as follows:
\begin{eqnarray}
(n u^\alpha)_{;\alpha} &=& 0 \ \ , \label{conti} \\
{T^{\alpha\beta}}_{;\beta} &=& -F^\alpha\ \ ,\label{dy}
\end{eqnarray}
where $F^\alpha = {T_{\rm rad}^{\alpha\beta}}_{;\beta}$ is the divergence of the
energy-momentum tensor $T_{\rm rad}$ for the photons, which gives the plasma's energy and momentum loss 
rate due to radiation. The cooling rate in the comoving frame of the flow is given by $\Lambda = -c
F^\alpha u_\alpha$.

For a steady flow, the continuity Equation (\ref{conti}) gives
\begin{equation}
\dot{M} = -4 \pi r^2 c\ u^r n\ m_p\ \ ,
\end{equation}
where the accretion rate $\dot{M}$ is a constant.

The law of local energy conservation is given by projecting Equation (\ref{dy}) onto
$u^\alpha$, for which
\begin{equation}
c\ u^\beta {{T_\beta}^\alpha}_{;\alpha} = \Lambda \ \ ,
\end{equation}
that is,
\begin{equation}
{\d\over \d\tau}\left({\epsilon\over n}\right) = -p{\d\over \d\tau}{1\over
n}+{\Gamma-\Lambda\over n}\ \ ,
\end{equation}
where
\begin{equation}
\Gamma = c\ u^\beta {{T_{\rm mag\ \beta}}^\alpha}_{;\alpha}=-{c\over 4\pi r}(2\ B^2\ u_r
+ r\ u_r\ B\ B^\prime)
\end{equation} 
is the heating term due to the annihilation of the magnetic field. As usual, we will use prime
to denote a derivative with respective to $r$ and an over-dot to denote a
derivative with respective to $t$. $T_{\rm mag}^{\alpha\beta}=(2 B^2 u^\alpha u^\beta + B^2
g^{\alpha\beta} - 2 B^\alpha B^\beta)/8\pi$ is the Maxwell stress-energy associated with the
frozen-in magnetic field.
 
The Euler equations are given by
\begin{equation}
(g^{\alpha\beta}+u^\alpha u^\beta){{T_\beta}^\gamma}_{;\gamma} = -F^\alpha+{\Lambda
\over c}u^\alpha \ \ .
\end{equation}
Only the radial component of the equation is non-trivial. It gives 
\begin{equation}
(p+\epsilon)u^r_{;\alpha}\ u^\alpha = -(g^{r\beta}+u^r\ u^\beta) p_{,\beta}\ \ ,
\end{equation}
where we have neglected the radiation pressure and magnetic pressure, which are
much smaller than the thermal pressure.

Then the equations for the radial velocity and temperature are:
\begin{eqnarray}
{1\over 2}{d\over dr}u_r^2&=&-{p^\prime\over p+\epsilon}\left(1+u_r^2-{2\ GM\over r
c^2}\right)-{GM\over r^2c^2}\ \ , \\
\alpha{T^\prime\over T} &=& {\Gamma-\Lambda\over k_b\ T\ n\ c\ u_r}-{1\over
\mu}\left({u_r^\prime \over u_r}+{2\over r}\right)\ \ .
\end{eqnarray}
Because the spacetime geometry is dominated by the supermassive black hole, all of these
equations ignore the effects of self-gravity within the accreting gas. These equations
are readily solved numerically to provide the radial structure of the infalling gas
once the boundary conditions and the magnetic field are specified.

\subsection{Boundary Conditions and Magnetic Field}

In principle, a precise knowledge of the wind sources surrounding the accretor would
provide all the necessary conditions at the capture radius to determine the structure
of the infalling gas uniquely.  However, only some global properties of the ambient
medium are known adequately well.  For example, at the Galactic center, the average 
density and flow velocity have been determined from emission-line measurements, but 
even there, a lack of information on the de-projected distance along the line of sight 
to the accretor prevents us from positioning the sources with certainty around the black 
hole.  The ambient conditions in the much more distant nucleus of M31 are even more
difficult to identify with precision. Our approach will instead be to use the 
accurately known spectrum produced by the accreting gas to guide our selection of
the boundary conditions, which therefore become adjustable parameters for the fit.

Taking our cue from the conditions prevalent at the Galactic center, we adopt a
wind velocity $v_\infty$ of about $500$ km s$^{-1}$, which then also determines 
the downstream plasma temperature for a strong shock:
\begin{equation}
T= (3\ v_\infty /4)^2\mu/3 R_g \ \ .
\end{equation}
Typically, $T\sim 3\times 10^6$ K. Three-dimensional hydrodynamic simulations show that
the region between the shocks and the transonic point is more or less isothermal. So we 
would expect the temperature at the outer boundary of our modeled region to 
be about several million degrees Kelvin.

HST observations (Lauer et al. 1998) show that the extended UV source in P2 has a radius of
about $0.2$ pc ($\sim 10^5 r_S$, where $r_S = 2GM/c^2$ is the Schwarszchild radius of
the central black hole). The outer boundary should have roughly the same scale as this. 
To ensure that this outer boundary is located below the transonic point (since we are
limiting our domain of solution to this region---see above), the radial velocity
of the flow must be larger than its thermal velocity at that point. We have already 
limited the temperature at the outer boundary. This therefore provides a lower limit for 
the radial velocity. At the same time, the flow must be bounded, so its radial
velocity cannot exceed the free fall value at this radius.

With $M$ and $u_r$ known, the accretion rate then depends on the ambient 
gas density. Although
Ciardullo et al. (1998) have argued that the ionized gas density in the nucleus of
M31 changes dramatically, dropping from $>10^4$ electrons cm$^{-3}$ at a radius of $\sim
7^{\prime\prime}$ or $23$ pc to $10^2$ electrons cm$^{-3}$ at $\sim 1^\prime$ or
$200$ pc, the actual density near the capture radius may still be quite different from 
this given the differences in length scales. As such, the accretion rate, or the gas
density at the outer boundary, remain as adjustable parameters.

The structure and intensity of the magnetic field are also poorly known in this
context, given the uncertain nature of magnetic field dissipation within a converging
flow.  Following the assessments of Kowalenko \& Melia (1999), 
and their application to Sgr A*
in the outer region of the accretion flow (Coker \& Melia 1997), we will invoke a
sub-equipartition field compared to the thermal energy density. That is, we will take
$B^2/8\pi=\beta_b\,\alpha\ n\ k_b\ T/2\mu$, where $\beta_b<1$. 

\subsection{Dichotomy of the Accretion Profiles}
\label{dich}

\begin{figure}
\figurenum{1}
\plotone{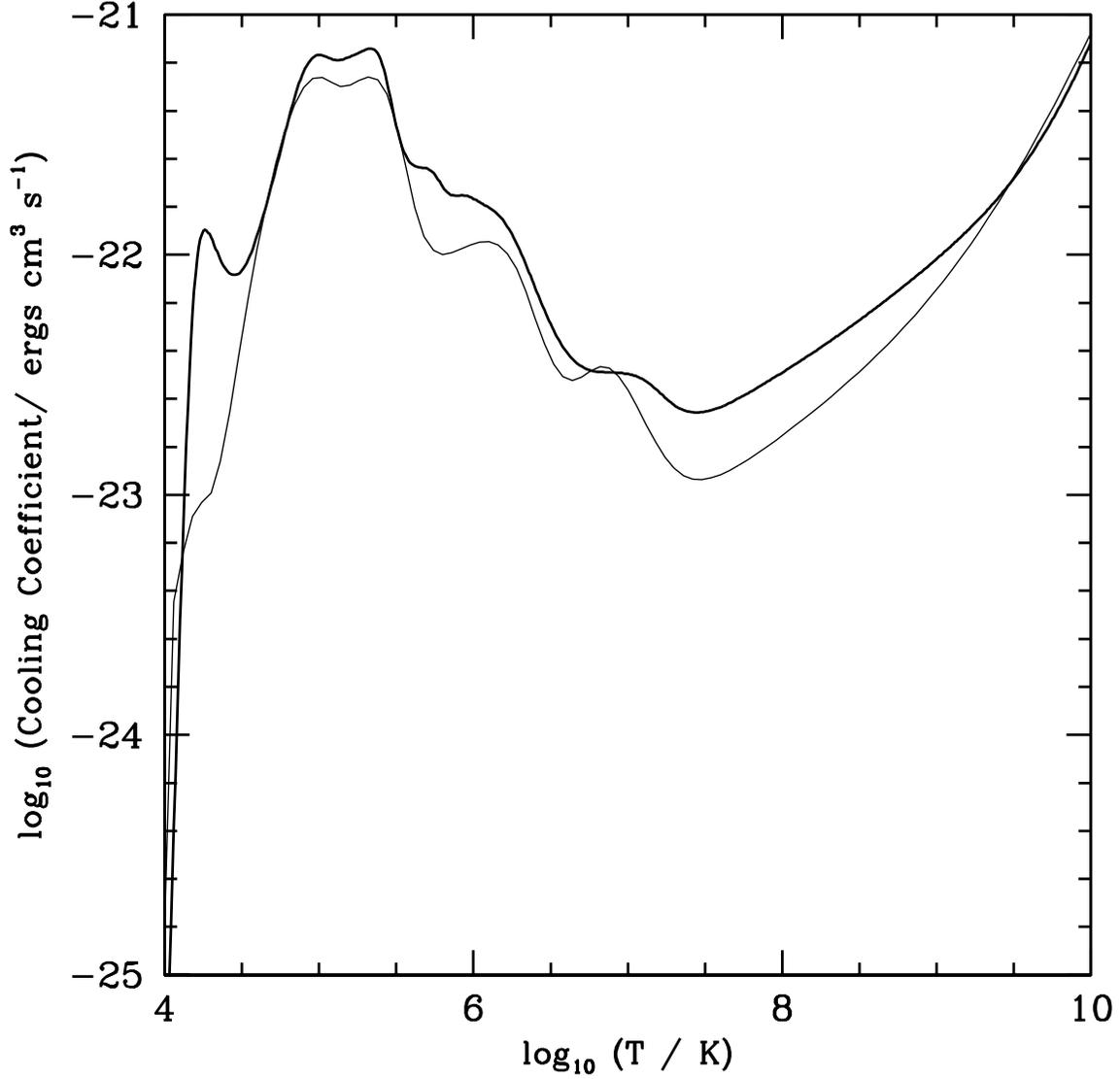}
\caption{Temperature dependence of the cooling coefficient, defined to be the cooling rate
divided by the baryon number density squared, for an optically thin plasma with cosmic
abundances. The thick solid curve is the coefficient used in this paper. The thin solid
curve is the result given by Gehrels \& Williams (1993), which we used in our previous
calculation (Liu \& Melia 2001).} 
\label{fig:cool}
\end{figure}

Figure~\ref{fig:cool} shows the cooling curve for an optically thin plasma with cosmic
abundances. The thin solid curve is that given by Gehrels \&
Williams (1993), which we used in our previous calculation. 
The thick solid curve gives a more
accurate cooling curve based on more detailed modeling with CLOUDY (Ferland 1996).  This is
the curve used for the calculations reported in this paper. It is evident that the plasma is
located in an unstable portion of this curve when its temperature lies between $10^5$ and
$10^7$ K.  As we have seen above, the temperature downstream of the capture radius is
typically several million Kelvin.  Thus, depending on the gas density in this region, which
determines the cooling rate at the outer boundary and the accretion rate for a given radial
velocity, the plasma may either heat up to about $10^{10}$ K, or cool down to around $10^4$ K.

\begin{figure}  
\figurenum{2}
\plotone{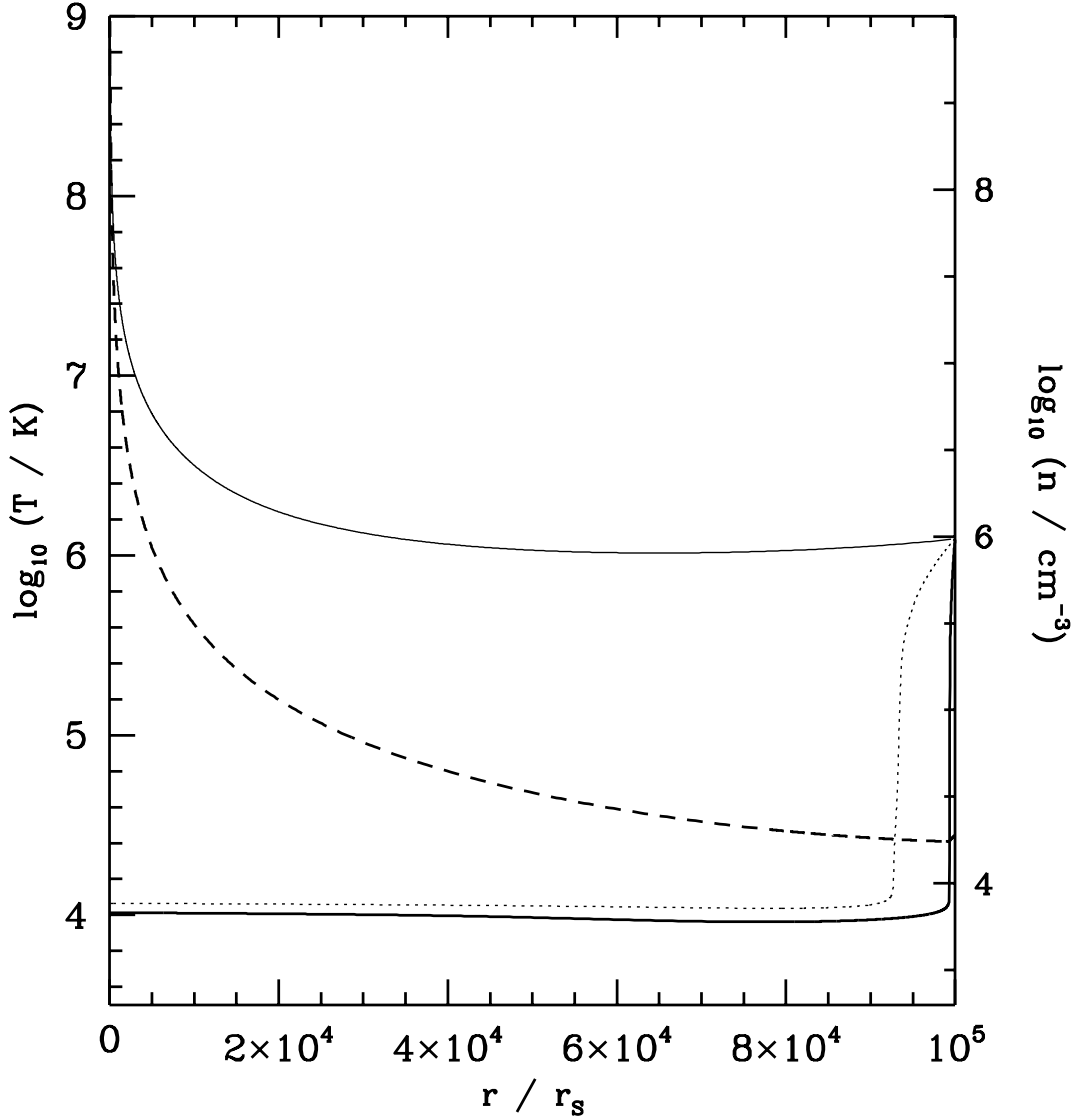}
\caption{Temperature profile for the accreting gas, for three values of accretion rate. The
initial temperature in all cases is $1.24\times 10^6$K .  The accretion rate are
$1.0\times 10^{23}$g s$^{-1}$ (thin, solid curve---the hot branch solution), $1.0\times
10^{24}$g s$^{-1}$ (dotted curve) and $1.0\times 10^{25}$g s$^{-1}$ (thick, solid curve---the
best fit model for M31*). Also shown here is the run of proton density (dashed curve) of the
best fit model, whose scale appears on the right hand side.}
\label{fig:profiles}
\end{figure}

Figure~\ref{fig:profiles} illustrates this dichotomy.  The thick solid curve corresponds to
our best fit model for the multi-wavelength observations of M31*, which has an outer boundary
$r_o$ of $10^5r_S$ ($\approx 0\farcs08$), a radial velocity of one half of the free fall
velocity, an accretion rate of $1.5\times 10^{25}$ g s$^{-1}$ and a magnetic parameter
$\beta_b=0.003$. The outer boundary temperature is given by the ratio of the thermal to
gravitational energy density, which is 0.068 for the best fit model. To emphasize the 
accretion rate dependence of the solutions, we also show in this figure the temperature
profiles for two configurations with different accretion rates, which are $1.5\times
10^{24}$ g s$^{-1}$ and $1.5\times 10^{23}$ g s$^{-1}$ for the dashed curve and the thin
solid curve, respectively. Clearly, the solutions depend rather sensitively on the accretion
rate. We will return to this point in section \ref{discuss} below.  It appears
that in order for M31* to simultaneously account for both the radio and UV
spectral components, the post-shock gas initiating its infall toward the central
accretor is highly unstable to cooling. For this reason, the temperature of the
plasma in the best-fit model drops quickly from $\sim 10^6$ K in the 
post-shock/capture region to $\sim 10^4$ K for the remainder of its inward flow. 

\begin{figure}
\figurenum{3} 
\plotone{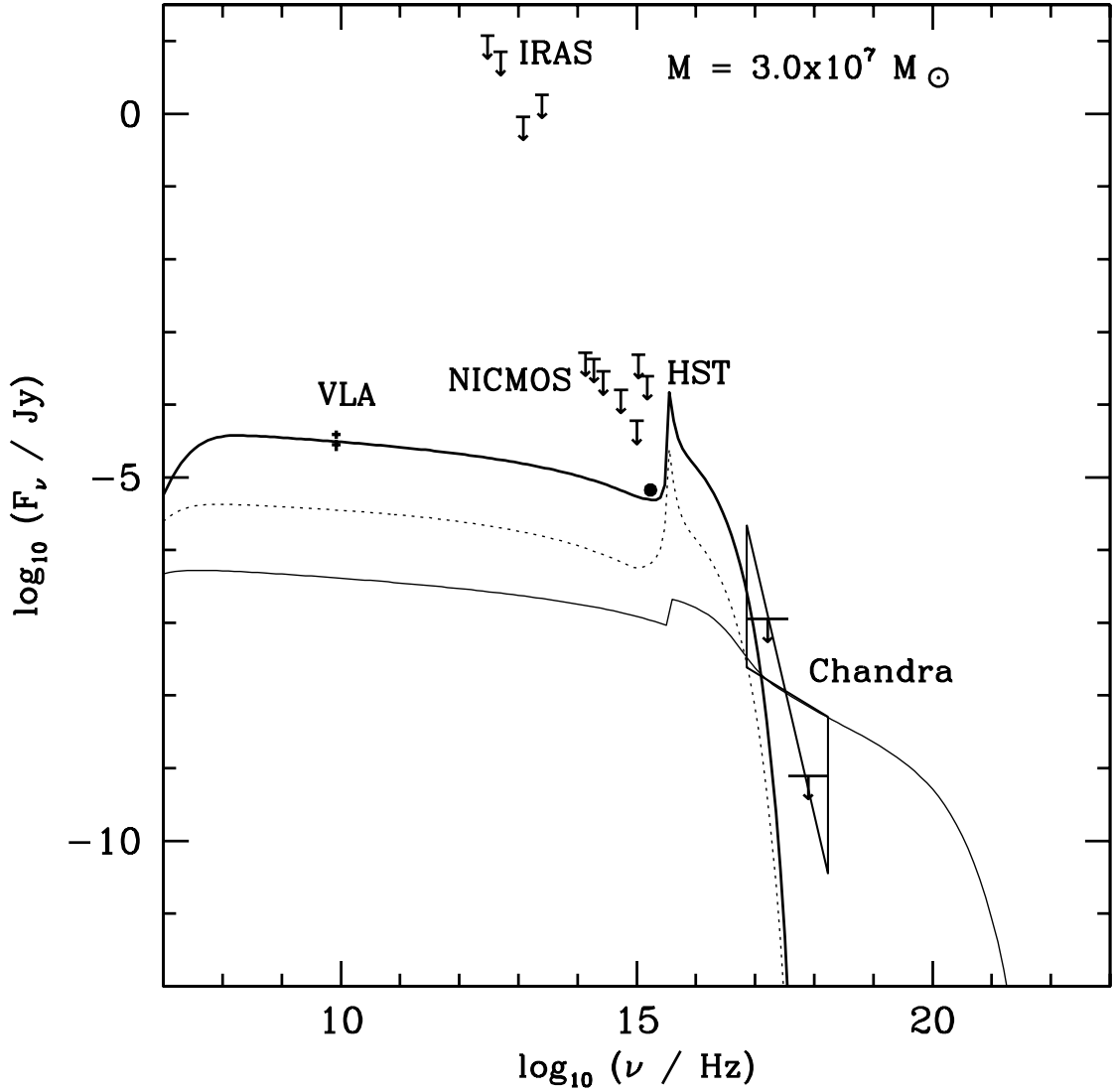}
\caption{A comparison of the overall spectrum arising from each of the three cases     
depicted in Fig.~\ref{fig:profiles}. The line types have the same definitions as
those in the figure. The X-ray upper limits correspond to the so-called ``southern"
source detected near M31* by {\it Chandra}, whereas the butterfly shows the
measured spectrum for the ``northern" source.}
\label{fig:solutions}
\end{figure}

In the hot branch, the cooling is dominated by thermal bremsstrahlung radiation. We would
expect a flat spectrum which cuts off at $\nu_c\sim k_b\ T/ h$. Figure~\ref{fig:solutions}
shows the spectra corresponding to the three configurations in Figure~\ref{fig:profiles}.  In
this figure, the two more constraining optical upper limits are from Lauer et al.  (1998) and
the three near Infrared upper limits are from Corbin et al (2001). The other data are
discussed in Liu \& Melia (2001). Note that these spectra do not include the 
contribution from cyclo-synchrotron emission, which is absent in the cool 
branch solutions, but would produce an additional peak at radio frequencies 
for the hot branch profiles. The magnitude and location of this cyclo-synchrotron 
peak would depend on the strength of the magnetic field.  Clearly,
the {\it Chandra} X-ray upper limit (or detection, depending on whether or not
one of the sources near M31* is its counterpart) already rules out the hot branch solution 
for M31*, under the assumption that the radio component extends into the UV region, 
even without the additional problems introduced at radio wavelengths by this 
additional component.  The cool branch solution, on the other hand, accounts very 
well for the radio and UV emission, and it features strong line emission due to the 
dominant line cooling component in the cooling function. Even so, the configuration 
with the low accretion rate can't produce the observed radio and UV flux, so there 
is a restricted range of parameter values at the outer boundary
that appear to be consistent with the data.

\section{CALCULATING THE EMISSION SPECTRUM}
\label{lines}

The emission spectrum of a cooling gas is a function of the density $n$ and temperature
$T$, which in turn are both functions of radius. With the assumption of spherical
symmetry, the mean emissivity (in units of energy per unit time per unit volume per unit
frequency interval) is 
\begin{equation}
4 \pi\, j_\nu(r,\theta) = 4 \pi\, \epsilon(r,\theta) \left\{ j_\nu^{\rm brem}(n(r),T(r)) +
\sum_{\rm lines}\ j^{\rm line}(n(r),T(r))\ \delta(\nu-\nu_{\rm line}(r,\theta)) \right\}
\ ,
\label{eq:j_nu}
\end{equation}
where
\begin{equation}
\epsilon(r,\theta) =  \left(\frac{\sqrt{1 - v(r)^2/c^2}}{(1 +
v(r)\cos{\theta}/c)}\right)^3 
\label{eq:eps}
\end{equation}
is the special relativistic correction to the monochromatic flux of moving emitters
(\cite{Corbin97}); $4 \pi\, j_\nu^{\rm brem}(n,T) \simeq 6.8 \times 10^{-38}\, n^2\, \Delta^2\,
T^{-1/2}\, \exp(-h\nu/k_bT)$~erg~s$^{-1}$~cm$^{-3}$~Hz$^{-1}$ is the mean emissivity due
to thermal bremsstrahlung (\cite{Frank92}), with the ionization fraction $\Delta=(1.0+0.445\,
T^{-1}e^{157890/T})^{-1}$ obtained following the prescription given in Rossi et
al. (1997); $j^{\rm line}(n,T)$ is the mean emissivity
(in units of energy per unit time per unit volume) due to line emission; and
\begin{equation}
\nu_{\rm line}(r,\theta) = \nu_{\rm line}^0 \left( \frac{1 - v(r) \cos{\theta}}{\sqrt{1
- v^2(r)}} \right) 
\label{nu}
\end{equation}
is the Doppler-corrected line frequency, with $\nu_{\rm line}^0$ being the rest-frame
line frequency. General relativistic effects have been ignored as the vast majority of
emission occurs at radii $r \gg r_S$.

Computationally, we view the inflow as a series of concentric spherical
shells, each of which is characterized by a set of physical parameters.
The velocity $v(r)$ and density $n(r)$ are specified by the equations of continuity and
motion, and the temperature $T(r)$ is specified by the energy balance equation,
including terms for line and bremsstrahlung cooling (\cite{Gehrels93}). 
The central mass is assumed to be $M = 3 \times 10^7 \ M_\odot$ and the mass accretion
rate is set at $\dot{M} = 1.5\times 10^{25}$g s$^{-1}$, corresponding to the best fit model
temperature profile shown in Figure~\ref{fig:profiles}.

We model each shell as a coronal equilibrium zone (appropriate for gas which is mainly
collisionally ionized) and use the photoionization code CLOUDY (\cite{Ferland96}; CLOUDY 
94.00) to compute the line emissivities.  Cosmic abundances are assumed 
throughout, with no grains present due to the high temperatures.
Radiative transfer effects between the zones are neglected, as gas is optically thin
above $0.1$GHz.

To obtain a spectrum of the flux density, the emissivity is integrated over the entire
flow volume:
\begin{equation}
F_\nu = \frac{1}{4 \pi\, D_{\rm M31}^2} \int 4 \pi\, j_\nu\ r^2\, \sin{\theta}\, dr\,
d\theta\, d\phi \ , 
\label{eq:F_nu}
\end{equation}
where $D_{\rm M31} = 784$ kpc is the distance to M31 (Stanek \& Garnavich 1998).
The flux density is then corrected for extinction along the line of sight due to
absorption (\cite{Cruddace74}; \cite{Morrison83}) and reddening (\cite{Cardelli89}),
assuming a column depth of $N_{\rm H} = 2.8 \times 10^{21}$~cm$^{-2}$ based on $Chandra$
observations (\cite{Garcia00}), cosmic elemental abundances (\cite{Dalgarno87}), $A_V =
0.24$ (\cite{Lauer93}), and $E(B-V) = 0.11$ (\cite{King95}).

\section{RESULTS}
\label{results}

\begin{figure}
\figurenum{4}
\rotatebox{270}{
\plotone{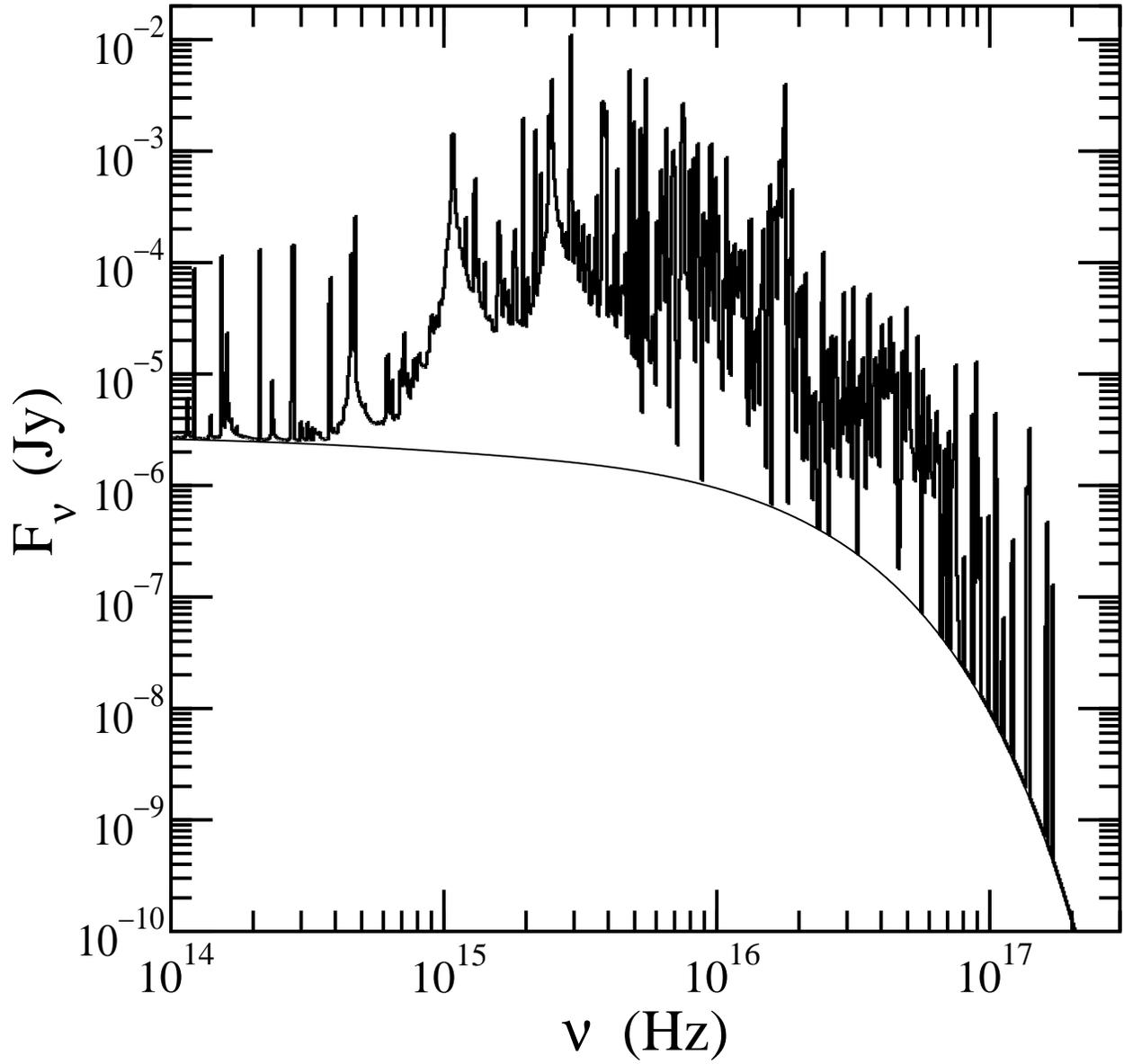}
}
\caption{Modeled intrinsic bremsstrahlung (thin curve) and bremsstrahlung plus line
(dark, heavy curve) spectra.}
\label{fig:intrinsic}
\end{figure}

Figure~\ref{fig:intrinsic} shows the intrinsic flux spectrum calculated with our model.
The spectrum (and cooling) is dominated by the UV recombination lines of warm Fe, O, Si,
and S species present in the outer flow region.

\begin{figure}
\figurenum{5}
\rotatebox{270}{
\plotone{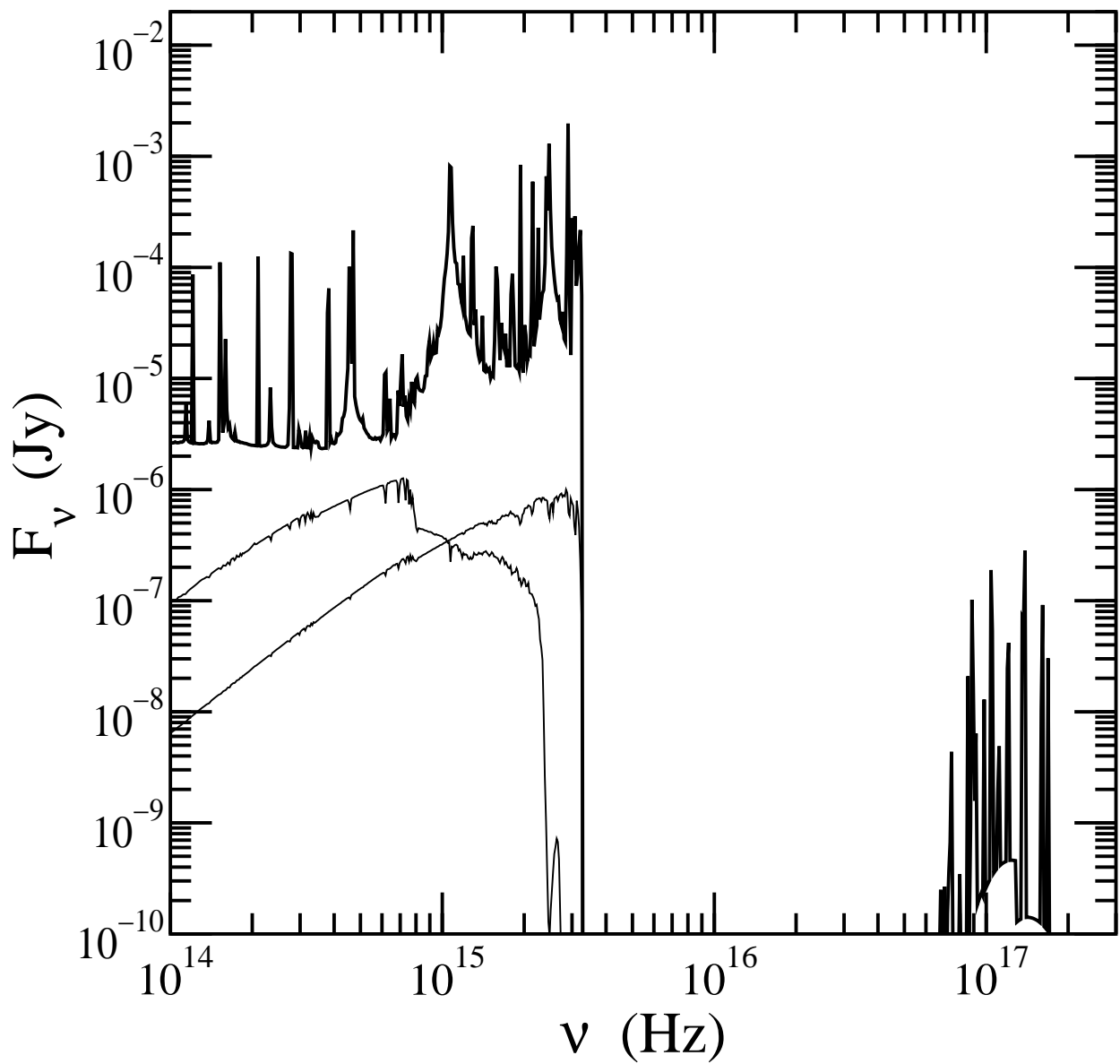}
}
\caption{Modeled line plus bremsstrahlung spectrum corrected for extinction and
reddening (dark).  Shown for comparison are stellar atmosphere spectra for $T =
10,000$~K and $T = 30,000$~K stars (thin curves), arbitrarily scaled.  See text for
description.}
\label{fig:corrected}
\end{figure}

Figure~\ref{fig:corrected} shows the model spectrum corrected for extinction.
It is evident that the strongest emission lines are effectively extinguished by the
intervening column of gas.
The surviving optical-UV and soft X-ray components both show very strong line emission
superimposed on a bremsstrahlung continuum, and should be easy to distinguish from,
e.g., the composite spectrum of a stellar population.
For comparison, the spectra of $T = 10,000$~K and $T = 30,000$~K stars computed from
Atlas stellar atmosphere modeling (\cite{Kurucz91}) are also presented in the figure.

\begin{figure}
\figurenum{6}
\rotatebox{270}{
\plotone{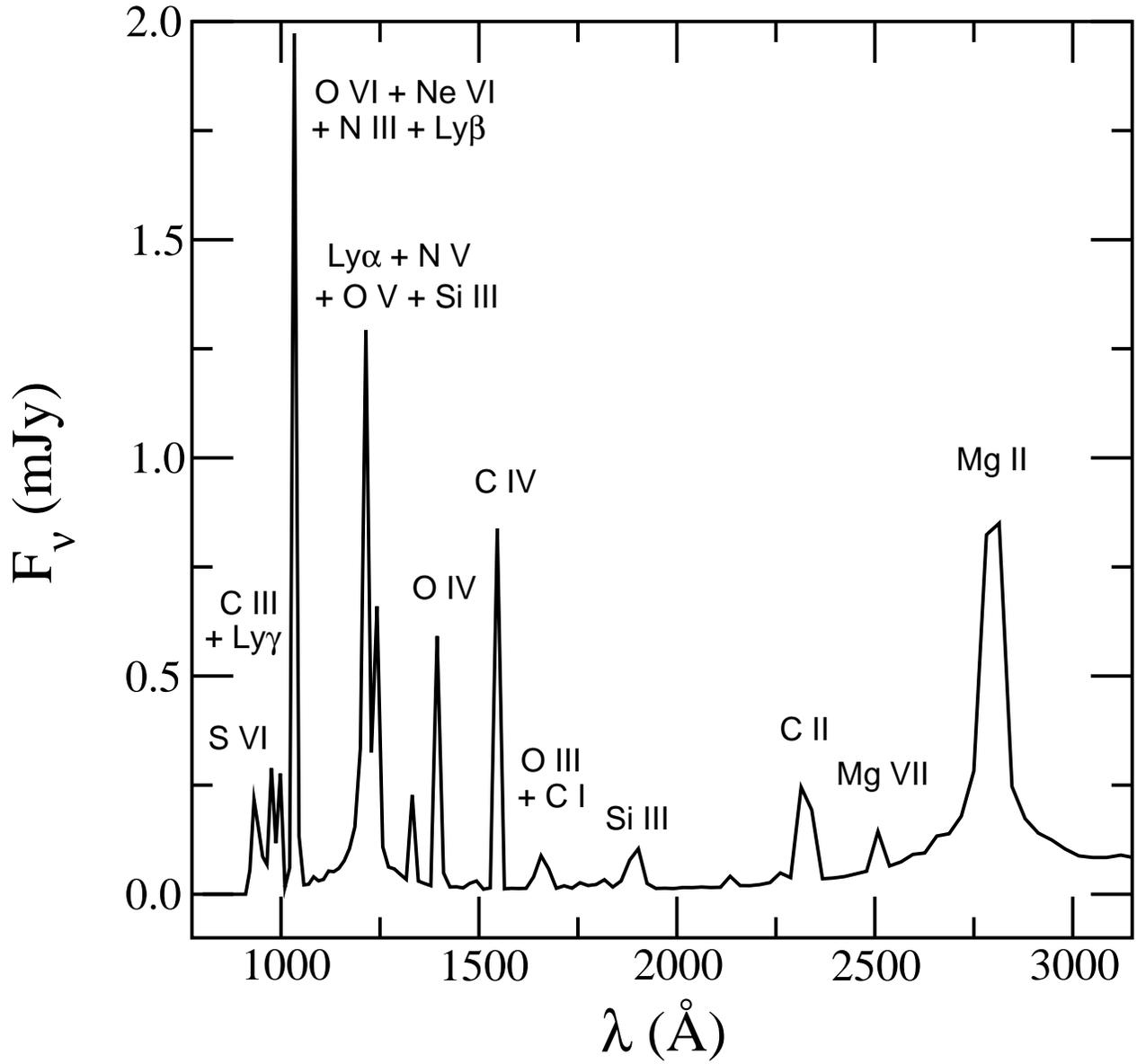}
}
\caption{Modeled ultraviolet spectrum.}
\label{fig:UV}
\end{figure}

\begin{figure}
\figurenum{7}
\rotatebox{270}{
\plotone{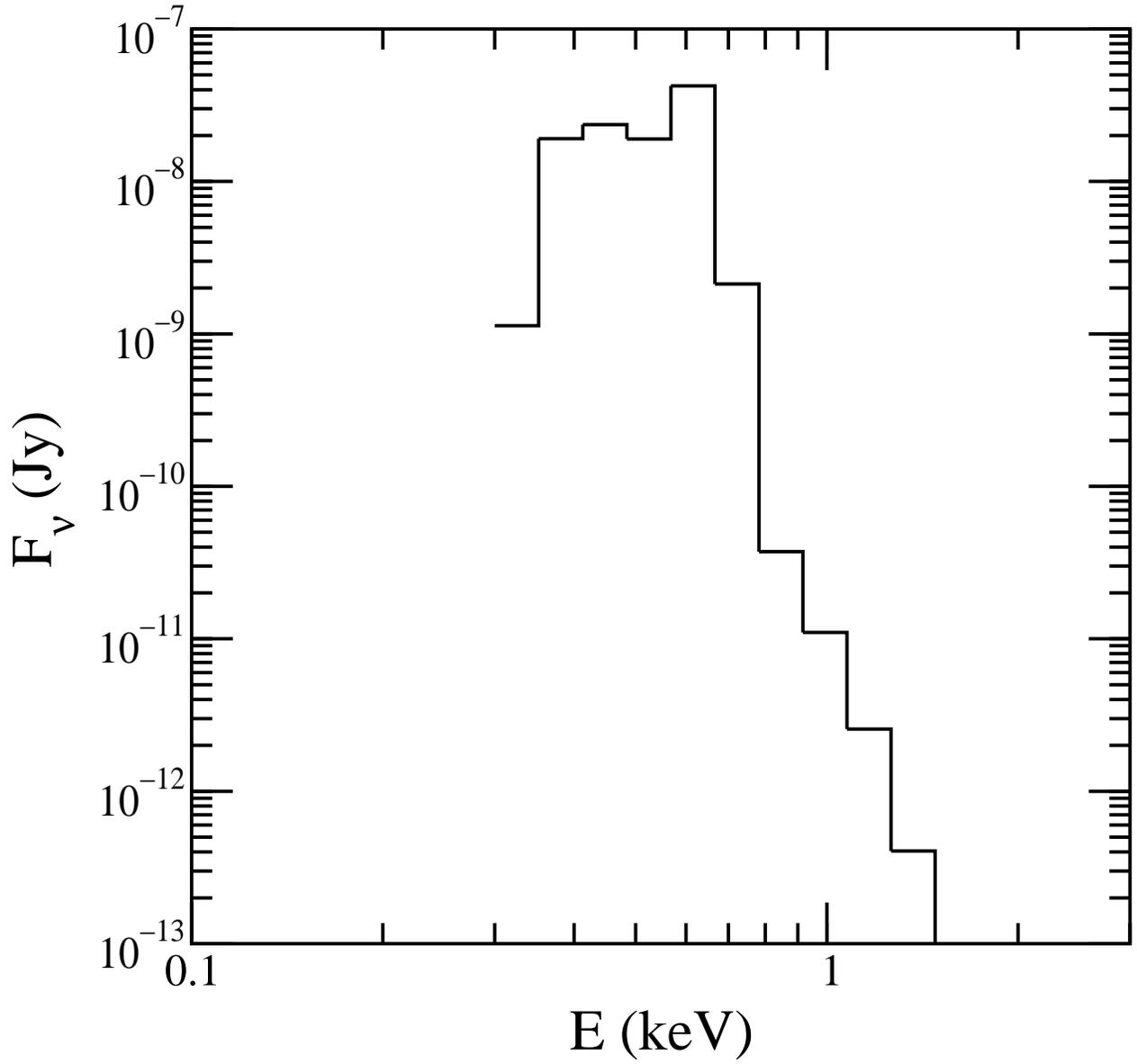}
}
\caption{Modeled X-ray spectrum, rebinned in equal logarithmic intervals.}
\label{fig:Xray}
\end{figure}

An expanded view of the modeled UV and X-ray spectra are shown in
Figures~\ref{fig:UV}~and~\ref{fig:Xray}. 
High spatial resolution observations of M31* at these wavelengths may provide a very
straight-forward test of this cooling accretion flow picture.
In particular, the UV spectrum is characterized by very strong line
emission; Table~\ref{tab:lines} lists the intensities of the strongest emission lines
relative to Ly${\alpha}$.

\section{DISCUSSION}
\label{discuss}

As mentioned in section \ref{dich}, these solutions are rather sensitive to the 
outer boundary conditions, particularly the accretion rate. For these simulations, 
we have used a simplified scheme in which compressional and magnetic heating balance 
line cooling in the outer region.  In reality, turbulent heating may also play an 
important role in stabilizing the temperature.
Additional heating may be provided by the disturbance generated by the stellar population
around P2. In any event, the essential feature of this model is that regardless of what
happens at large radii, line cooling can dominate over the heating processes at small radii
when the accretion rate is large enough. This inner cool region is important because the
observed weak radio emission does not allow the plasma to be heated up. The two radio
measurements of M31* give a flux density of $28-39\ \mu$Jy at $3.6\cm$ (Crane et al. 1992 \&
1993). The $3.6\cm$ radio flux from an optically thick spherical emitter can be estimated
using the following equation: \begin{eqnarray} F(3.6\cm)&=&\left({r_S\over
D_{M31}}\cdot{R\over r_S}\right)^2\ {2\pi\,k_b\,T\over \lambda^2} \nonumber \\ &=& 
0.895\,{T\over 10^{10}\K}\left({R\over r_S}\right)^2\ \mu{\rm Jy}\ . \end{eqnarray} 
Thus, a source with radius $10\ r_S$ and a temperature of several times $10^{10}$ K 
will produce far too much flux at $3.6\cm$. On the other hand, a hot optically thin 
source cannot produce the strong UV emission
via self-Comptonization of the radio photons. If the UV component is in fact
associated with M31*, this argument also rules out many other models
developed in recent years to account for the emissivity of supermassive black holes in the
cores of nearby galaxies, including Sgr A* at the Galactic center (the ADAF model, Narayan,
Yi, \& Mahadevan 1995; the ADIOS model, Blanford \& Begelman 1999; and the jet model, Falcke
\& Markoff 2000). However, if the UV emission is not associated with M31*, 
a hot optically thin plasma may still account for the radio emission, as it does
for Sgr~A*.  Distinguishing between these two scenarios via the observation
of UV lines in M31*'s spectrum makes our proposed measurements highly desirable.

\begin{figure}
\figurenum{8}
\plotone{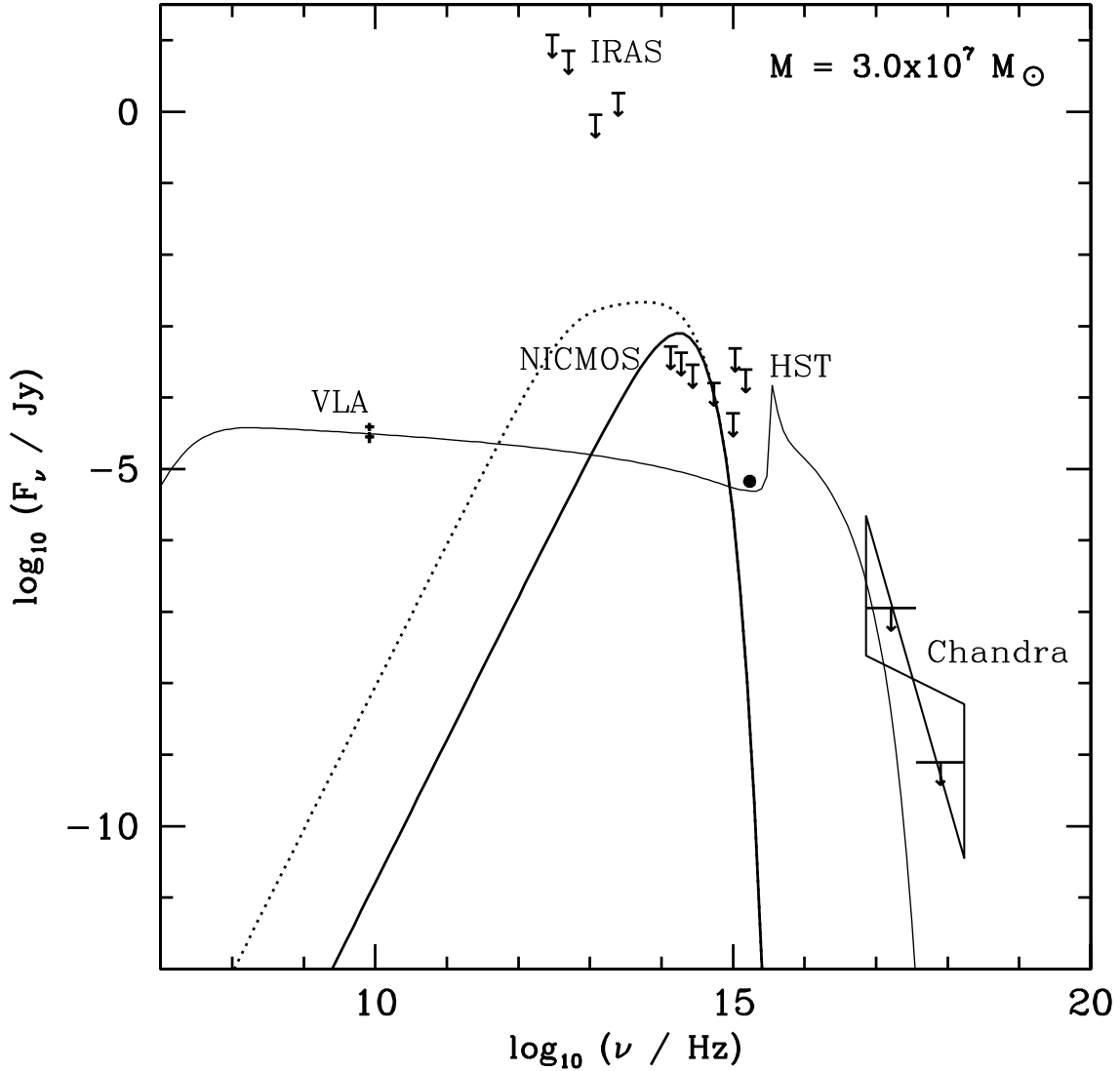}
\caption{Spectra of optically thick disks. The disks have an inclination angle of
$60^\circ$, an inner boundary of $3 r_S$ and the accretion rate is $5.0\times 10^{18}$ g
s$^{-1}$. The thick solid
line corresponds to a small disk with an outer boundary of $10 r_S$, while the dotted line
corresponds to a disk with an outer boundary of $1000 r_S$. The thin solid line corresponds
to the best fit radial accretion model, though here only the hydrogen line emission is 
included. As in Fig. 3, the X-ray upper limits correspond to the so-called ``southern"
source detected near M31* by {\it Chandra}, whereas the butterfly shows the
measured spectrum for the ``northern" source. In this best fit, we have treated these
X-ray data as upper limits to the actual spectrum of M31* (see text).} 
\label{fig:disk}
\end{figure}

In calculating the spectrum for the hot branch solution in Figure~\ref{fig:solutions}, 
we have
not included the effects due to a magnetic dynamo, should the gas circularize before it
reaches the event horizon.  In the case of Sgr A*, the dominant contribution to the mm and
sub-mm spectrum is made by a Keplerian structure within the inner $5\ r_S$ or so (Melia, Liu
\& Coker 2000, 2001). In the case of M31*, this region may or may not exist, depending on how
much specific angular momentum is carried inward.  For the same conditions as in Sgr A*, M31*
would not have a Keplerian region since $r_S ({\rm M31*})\approx 
10\ r_S({\rm Sgr A*})$, i.e., the gas would not circularize before crossing 
the event horizon. So the model for Sgr A* may not work here. The much larger 
accretion rate implied by our model also rules out the possibility of a standard
optically thick disk being present in M31*.  As shown in Figure~\ref{fig:disk}, even with an
accretion rate more than $6$ orders of magnitude smaller than the accretion rate we infer for
the best fit case, a small optically thick disk still produces too much optical radiation
compared to the observations.  Such an optically thick disk would not be able to account for
the UV upturn either, making it unlikely that such a structure is present in the nucleus of
M31.

In Figures 3 and 8, we have treated the {\it Chandra} flux measurements of the two
nearest X-ray sources to M31* as upper limits, since neither of these has been
established with confidence as the actual counterpart to the radio source.
A slightly different fit, assuming that the so-called ``southern'' source is indeed
the counterpart, was presented in Liu \& Melia (2001). None of the conclusions
presented in this paper are affected qualitatively in this instance, except for
slight differences in the recombination line intensities.  However, if the
``northern'' source turns out to be the actual counterpart (see Garcia et al. 2000),
the cold accretion model will not work for M31*.  A close inspection of Figure 8
shows that the X-ray spectrum (indicated by the butterfly) is then too hard to
be described adequately by bremsstrahlung emission.  The nature of M31* is therefore
still rather uncertain, making further study of this very interesting object
highly desirable.

\section{ACKNOWLEDGMENTS}

This work was supported by an NSF Graduate Fellowship and by NASA grants NAG5-8239 and
NAG5-9205 at the University of Arizona, and by a Sir Thomas Lyle Fellowship and a
Miegunyah Fellowship for distinguished overseas visitors at the University of Melbourne.

\clearpage
 
\begin{deluxetable}{lc}
\footnotesize
\tablecaption{Attenuated Emission Line Intensities\label{tab:lines}}
\tablewidth{0pt}
\tablehead{\colhead{Line\tablenotemark{\dagger}} &
\colhead{Intensity\tablenotemark{\ddagger}}}

\startdata
Ly${\alpha}$~$\lambda 1216$ & 1.00 \nl
\ion{O}{6} $\lambda\lambda 1032,1038$ & 0.59 \nl
\ion{Mg}{2} $\lambda\lambda 2796,2804$ & 0.47 \nl
\ion{C}{4} $\lambda\lambda 1548,1551$ & 0.16 \nl
\ion{N}{5} $\lambda\lambda 1239,1242$ & 0.13 \nl
\ion{O}{4} $\lambda 1399$ & 0.13 \nl
\ion{S}{6} $\lambda\lambda 933,945$ & 0.07 \nl
\ion{C}{2} $\lambda 2326$ & 0.06 \nl
\ion{C}{3} $\lambda 977$ & 0.06 \nl
\ion{Ne}{6} $\lambda 999$ & 0.06 \nl
\ion{O}{5} $\lambda 1215$ & 0.03 \nl
\ion{Si}{3} $\lambda 1207$ & 0.03 \nl
\ion{Si}{3} $\lambda 1891$ & 0.03 \nl
Ly${\beta}$~$\lambda 1026$ & 0.02 \nl
Ly${\gamma}$~$\lambda 973$ & 0.02 \nl
\ion{C}{1} $\lambda 1657$ & 0.02 \nl
\ion{N}{3} $\lambda 991$ & 0.02 \nl
\ion{O}{3} $\lambda\lambda 1661,1666$ & 0.02 \nl
\enddata

\tablenotetext{\dagger}{All wavelengths in \AA.}
\tablenotetext{\ddagger}{Relative to Ly${\alpha}$~$\lambda 1216$.}

\end{deluxetable}
\clearpage

\clearpage

\clearpage

\clearpage

\clearpage

\clearpage

\clearpage

\clearpage

\begin{thebibliography}{DUM}

%\bibitem[Aitken et al. 2000]{Ait00}{Aitken, D.K., Greaves, J.,                                 
%Chrysostomou, A., Jenness, T., Holland, W., Hough, J.H., Pierce-Price,                         
%D. \& Richer, J. 2000, ApJ Letters, 534, L173}                                                 
                                                                                               
%\bibitem[Baganoff et al. 2001]{Bag01}{Baganoff, F. et al. 2001, ApJ submitted}                 
                                                                                               
\bibitem[Blandford \& Begelman 1999]{BB99}{Blandford, R.D. \& Begelman,                        
M.C. 1999, MNRAS, 303, L1}                                                                     
                                                                                               
\bibitem[Brown et al. 1998]{Bro98}{Brown, T.M., Ferguson, H.C., Stanford, S.A.                 
\& Deharveng, J.-M. 1998, ApJ, 504, 113}                                                       
                                                                                               
%\bibitem[Brown \& Mathews 1970]{BM70}{Brown, R.L. \& Mathews, W.G., 1970,                      
%ApJ, 160, 939}                                                                                 
                                                                                            
\bibitem[Cardelli, Clayton, \& Mathis 1989] {Cardelli89} Cardelli, J. A.,  Clayton,
G. C., \& Mathis, J. S. 1989, \apj, 345, 245 

\bibitem[Chandrasekhar 1939] {chandra39}  Chandrasekhar, S. 1939. An Introduction to the
Study of Stellar Structure (New York: Dover)
                                                                                               
\bibitem[Ciardullo et al. 1988]{Ci88}{Ciardullo, R., Rubin, V.C.,                              
Jacoby, G.H., Ford, H.C. \& Ford, W.K. Jr.  1988, AJ, 95, 438}                                 

\bibitem[Coker 1997] {Coker97} Coker, R.F. \& Melia, F. 1997, ApJ Letters, 488, L149

%\bibitem[Coker 1999]{Coker99} Coker,R.F. 1999, Thesis(PHD)

\bibitem[Coker \& Melia 2000]{CM00}{Coker, R.F. \& Melia, F. 2000, ApJ,                        
534, 723}                                                                                      

\bibitem[Corbin 1997] {Corbin97} Corbin, M. R. 1997, \apj, 485, 517 

\bibitem[Corbin 2001] {Corbin01} Corbin, M.R., O'neil, E. \& Reike, M.J. 2001, \aj,
submitted

\bibitem[Crane et al. 1992]{Cr92}{Crane, P.C., Dickel, J.R. \& Cowan, J.J. 1992,               
ApJ Letters, 390, L9}                                                                          
                                                                                               
\bibitem[Crane et al. 1993]{Cr93}{Crane, P.C., Cowan, J.J., Dickel, J.R. \&                    
Roberts, D.A. 1993, ApJ Letters, 417, L61}                                                     

\bibitem[Cruddace et al. 1974] {Cruddace74} Cruddace, R., Paresce, F., Bowyer, S., \&
Lampton, M. 1974, \apj, 187, 497

\bibitem[Dalgarno \& Layzer 1987] {Dalgarno87} Dalgarno, A., \& Layzer, D. 1987,
Spectroscopy of Astrophysical Plasmas (Cambridge: Cambridge University Press)

\bibitem[Falcke \& Markoff 2000]{FM00}{Falcke, H. \& Markoff, S. 2000,                         
A\&A, 362, 113}                                                                                

\bibitem[Ferland 1996] {Ferland96} Ferland, G. J. 1996, Hazy, a brief introduction to
CLOUDY 94.00 

\bibitem[Frank, King, \& Raine 1992] {Frank92} Frank, J., King, A., \& Raine, D. 1992,
Accretion Power in Astrophysics (Cambridge: Cambridge University Press)

\bibitem[Garcia et al. 2000] {Garcia00} Garcia, M. R., Murray, S. S., Primini, F. A.,
Forman, W. R., Jones, C., McClintock, J. E., \& Jones, C. 2000, \apj, 537, L23

\bibitem[Garcia et al. 2001]{Gar01}{Garcia, M.R., Murray, S.S., Primini, F.A.,                 
Forman, W.R., Jones, C., \& McClintock, J.E. 2001, IAU 205, Galaxies at                        
the Highest Angular Resolution, eds. Schilizzi, Vogel, Paresce, \& Elvis, in press.}           

\bibitem[Garcia 2001]{Gar}{Garcia, M. R., 2001, private communication.}

\bibitem[Gehrels \& Williams 1993] {Gehrels93} Gehrels, N. \& Williams, E. D. 1993,
\apj, 418, L25

\bibitem[King, Stanford, \& Crane 1995] {King95} King, I. R., Stanford, S. A., \& Crane,
P. 1995, \aj, 109, 164

\bibitem[Kormendy et al. 1999]{Kor99}{Kormendy, J. \& Bender, R. 1999, ApJ, 522, 772}          
                                                                                               
\bibitem[Kowalenko \& Melia 1999]{KM99}{Kowalenko, V. \& Melia, F. 1999,                       
MNRAS, 310, 1053}                                                                              
                                                                                               
\bibitem[Kurucz 1991] {Kurucz91} Kurucz, R. L. 1991, Proceedings of the Workshop on
Precision Photometry: Astrophysics of the Galaxy, eds. Davis, Philip, Upgren, \& James
(Davis: Schenectady)

\bibitem[Lauer et al. 1993] {Lauer93} Lauer, T. R., et al. 1993, \aj, 106, 1436

\bibitem[Lauer et al. 1998] {Lauer98} Lauer, T.R., Faber, S.M., Ajhar, E.A., Grillmair,
C.J. \& Scowen, P.A. 1998, \aj, 116, 2263

\bibitem[Liu \& Melia 2001] {Liu01} Liu, S. \& Melia, F. 2001, ApJ Letters, 550, L151

\bibitem[Melia 1992]{Me92}{Melia, F. 1992, ApJ Letters, 398, L95}                             

\bibitem[Melia 1994]{Me94}{Melia, F. 1994, ApJ, 426 577}
                                                                                               
\bibitem[Melia, Liu \& Coker 2001]{MLC00a}{Melia, F., Liu, S. \& Coker,                       
R.F. 2001, ApJ, in press}                                                                     
                                                                                               
\bibitem[Melia, Liu \& Coker 2000]{MLC00b}{Melia, F., Liu, S. \& Coker,                       
R.F. 2000, ApJ Letters, 545, L117}                                                            
                                                                                               
\bibitem[Morrison \& McCammon 1983] {Morrison83} Morrison, R., \& McCammon, D. 1983,
\apj, 270, 119
                                                                                               
\bibitem[Narayan, Yi, \& Mahadevan 1995]{NYM95}{Narayan, R., Yi, I.                            
\& Mahadevan, R. 1995, Nature, 374, 623}                                                       
                                                                                               
\bibitem[Novikov \& Thorne 1973] {Novikov} Novikov, I.D. \& Thorne, K.S. 1973, Black
Holes (edited by C.DeWitt \& B.S.DeWitt, Gordon and Breach Science Publishers)

\bibitem[Rossi 1997]{Rossi} {Rossi, P., Bodo, G., Massaglia, S., \& Ferrari, A. 1997, AA,
321,672}
                                                                                               
\bibitem[Shapiro]{Sh1972}{Shapiro, S.L. 1973a, ApJ, 180, 531}                                   

\bibitem[Shapiro 1973]{Shapiro} Shapiro, S.L. 1973b, \apj, 185, 69
                                                                                              
\bibitem[Stanek et al. 1998]{Stan98}{Stanek, K.Z. \& Garnavich, P.M. 1998 ApJ Letters,
503, L131}                                                                                          
                                                                                               
%\bibitem[Sutherland]{Suth93}{Sutherland, R.S. \& Dopita, M.A., 1993, ApJS, 88, 253}

\bibitem[Tremaine 1995]{Tre95}{Tremaine, S. 1995, AJ, 110, 628}                                
                                                                                               
\bibitem[Trinchieri]{TF91}{Trinchieri, G. \& Fabbiano, G. 1991, ApJ, 382, 82}                  
                                                                                              
\end{thebibliography}
\end{document}